\documentclass[11pt]{article}
\usepackage{amsmath,amsfonts,amsthm}
\author{Sergey I. Nikolenko \thanks{St. Petersburg State University, Department of Mathematics and Mechanics, St. Petersburg, Russia. E-mail: {\tt smartnik@inbox.ru} } }
\title{Hard satisfiable instances for DPLL-type algorithms}

\newcommand{\n}[1]{\overline{#1}}
\newtheorem{thm}{Theorem}

\begin{document}
\maketitle
\section{Introduction}
Satisfiability is one of the most popular NP-complete problems.
There are two main types of algorithms for solving SAT, namely
local search (for references see, for example, \cite{DHIV}) and
DPLL-type (this type was first described in the work \cite{DP} of
Davis and Putnam and \cite{DLL} of Davis, Logemann and Loveland).
A lot of effort has been invested in proving "less-that-$2^N$"
upper bounds for such algorithms. In this paper we concentrate on
proving exponential lower bounds and consider two DPLL-type
algorithms: GUC (Generalized Unit Clause heuristic; introduced in
\cite{ChaoFr}) and Randomized GUC.

DPLL-type algorithms were historically the first
``less-than-$2^N$'' algorithms for SAT. They receive as input a
formula $F$ in CNF with variables $x_1,\ldots,x_N$. After that, a
DPLL-type algorithm simplifies the input according to a certain
set of \emph{transformation rules}. If the answer now is obvious
(the simplified formula is either empty or contains a pair of
contradicting unit clauses), the algorithm returns an answer. In
the opposite case, it chooses a literal $l$ in the formula
according to a certain heuristic. Then it constructs two formulas,
one corresponding to $l:=true$ and the other to $l:=false$, and
recursively calls itself for these two formulas (note that since
we deal with the running time of the algorithm, the order in which
it calls itself for these two formulas does matter).  If any of
the calls returns the answer ``Satisfiable'', the algorithm also
returns this answer. Otherwise, it returns ``Unsatisfiable''.
Therefore, such algorithms differ from each other by two
procedures: one for simplifying a formula, and the other for
choosing the next literal.

Superpolynomial lower bounds for regular resolution (and hence
DPLL-type algorithms) are known since \cite{Tseit}. In
\cite{Achli}, a probabilistic distribution of exponentially hard
formulas was considered, and an exponential lower bound for two
DPLL-type algorithms (GUC and UC) was proved. Supposedly, among
them satisfiable instances should exist, but this question remains
open. Exponentially hard satisfiable instances were found for
local search algorithms (see \cite{HirLS}). However, no results
have been published about exponentially hard \emph{provably
satisfiable} formulas for DPLL-type algorithms. In this paper we
present such instances. Section 2 is devoted to basic definitions,
in section 3 examples of hard unsatisfiable formulas are given,
sections 4 and 5 contain proofs of lower bounds for the
algorithms, and open questions are formulated in Section 6.

\section{Preliminaries}
We denote by $X$ a set of boolean variables. The negation of a
variable $x$ is denoted by $\n{x}$. If $U\subseteq X$, then
$\n{U}=\{\n{x}\mid x\in U\}$. {\it Literals} are members of the
set $X\cup\n{X}$. A {\it clause} is a set of literals that does
not contain simultaneously any variable together with its
negation. A {\it formula in CNF} is a finite set of clauses. A
clause is called {\it unit} if it consists of one literal. A
literal is called {\it pure} with respect to a formula if the
formula contains only the literal, but does not contain its
negation. We denote by $PL(F)$ the collection of all pure literals
in $F$.

An {\it assignment} is a finite subset $I\subseteq X\cup\n{X}$
that does not contain any variable together with its negation. We
denote by $F[I]$ a formula that results from $F$ and an assignment
$I=\{x_1,x_2,\ldots,x_n\}$ after removing all clauses containing
the literals $x_i$ and deleting all occurrences of the literals
$\n{x_i}$ from the other clauses. An assignment $I$ is said to
{\it satisfy} the formula $F$, if $F[I]$ is the empty formula
(that is, $F[I]$ contains no clauses).

For a formula $F(x_1,\ldots,x_n)$ we construct its binary {\it
assignment tree}. Its nodes are partial assignments for $F$
consisting of literals $x_1,\ldots,x_n$ or their negations, and
the sons of a node $I=\{l_1,\ldots,l_i\}$, where
$l_j\in\{x_j,\n{x_j}\}$, are the assignments
$I_1=\{l_1,\ldots,l_i,x_{i+1}\}$ and
$I_2=\{l_1,\ldots,l_i,\n{x_{i+1}}\}$. Following \cite{ChaoFr}, we
denote by $C^F_i$ the collection of clauses in $F$ containing
exactly $i$ literals (we will omit the upper index if it is clear
from context).

\section{Hard unsatisfiable formulas}
First examples of unsatisfiable formulas requiring superpolynomial
time for regular resolution (shown in \cite{Galil} to be
equivalent to the Davis-Putnam procedure for complexity issues)
appeared in \cite{Tseit}. Examples in that article were obtained
by using boolean formulas based on graphs. Tseitin used rather
simple graphs, and his bounds were improved by Galil in
\cite{Galil}. In \cite{Urqu}, using the graph theory results on
expanders, the bounds were improved to the form of $2^{cN}$, where
$N$ is the number of variables in the formula (in the future $c$
will denote that very constant). Note that the bounds proven in
the present article depend on the best known bound for
unsatisfiable formulas and, therefore, will automatically improve
if the above-mentioned constant $c$ is increased.

Let us quote the following theorem from \cite{Urqu} (here $S_m$ is
an always existing, previously constructed in the same article
formula):
\begin{thm}[\cite{Urqu}, 5.7]
There is a constant $c>1$ such that for sufficiently large $m$,
any resolution refutation of $S_m$ contains $c^n$ distinct
clauses, where $S_m$ is of length $O(n), n=m^2$.
\end{thm}

In \cite{PudImp}, using a generalization of Tseitin's tautologies,
the following result was established: for every $k\ge3$, there
exists a constant $c_k>0$, $c_k=O(1/k^{1/8})$, such that every
DPLL-algorithm for $k$-SAT has worst-case time complexity at least
$\Omega(2^{N(1-c_k)})$, where $N$ is the number of variables in
the formula.

It is also worth mentioning that formulas in \cite{PudImp} and
\cite{Urqu} have linear number of clauses, that is, there is a
constant $b$ such that these formulas have less than $bN$ clauses,
where $N$ is the number of variables in them.

We denote by $G_k(y_1,\ldots,y_N)$ the hard formula in $k$-CNF
appearing in \cite{PudImp} with $N$ variables $y_1,\ldots,y_N$.

\section{Hard formulas for GUC}
\begin{figure}
\fbox{\parbox{12 cm}{
\parskip=-2mm
{\bf Making a choice with the GUC algorithm\\ Input:} A formula $F$ in CNF.\\
{\bf Method:}
\begin{enumerate}
\parskip=-1mm
\item $m:=min\{i:\ C_i^F\ne\emptyset\}$. \item If $m=1$ then
choose $C$ randomly from $C_1^F$ and set $l$ to the only literal
in $C$.

\item Else if $PL(F)\ne\emptyset$, choose $l$ randomly from
$PL(F)$.

\item Else choose $C$ randomly from $C_m^F$, then choose $l$
randomly from $C$.

\item Output $l$.
\end{enumerate}
}} \caption{One step of the improved GUC algorithm}\label{GUC}
\end{figure}

The GUC algorithm is described in \cite{ChaoFr} and its procedure
for making a choice is shown here on Fig.\ref{GUC}. Essentially,
it selects a random literal satisfying a clause of the smallest
size. Compared to the algorithm in \cite{ChaoFr}, we have added
the pure literals rule to its choice heuristic, that is, if the
negation of a literal does not occur in the formula, we
automatically satisfy this literal. Obviously, checking for pure
literals can be done in polynomial number of steps (with respect
to the number of variables). It is also obvious that applying the
pure literals rule cannot make the current partial assignment
contradictory. Note that our bounds also hold for the original GUC
algorithm (and, later, the original Randomized GUC algorithms
instead of the modified version); in fact, the pure literals rule
will sometime make our bounds worse.

In this article we use the backtracking implementation described,
for example, in \cite{Achli}. Basically, every time an algorithm
splits on some variable, it makes a choice, and the number of such
choices in that case measures its efficiency. When first reaching
a node, algorithm marks a choice it has to make as {\it forced},
if it was made by using the transfomation rules. In our case, such
choices occur when there is either a unit clause or a pure literal
in the formula. In the opposite case we will call a choice {\it
free}. The backtracking implementation of GUC will ``go down the
assignment tree'' until it finds a contradiction, and then
backtrack to the last free choice. Then it flips the value
assigned during this last free choice, marks this choice as
forced, and continues. We measure the complexity of our algorithm
as the number of choices (both free and forced) it makes until it
finds a satisfying assignment.

Let us now proceed to proving the exponential lower bound on
satisfiable formulas. Consider the following formula (we denote by
$x\vee E$ the set of clauses obtained by adding $x$ to all clauses
in $E$):
\begin{multline*}
F=(x_1\vee G_k(x_{M+1},\ldots,x_{\lceil
\frac{1+c_k}{c_k}M\rceil}))\wedge\\\wedge(\n{x_1}\vee
x_2\vee\n{x_3})\wedge(\n{x_1}\vee
x_3\vee\n{x_4})\wedge\cdots\wedge(\n{x_1}\vee
x_{M-1}\vee\n{x_M})\wedge(\n{x_1}\vee x_M\vee\n{x_2})
\end{multline*}
Note that the second line corresponds to $\n{x_1}\lor H$, where
$H$ is a formula forcing the variables $x_2,\ldots,x_M$ to have
equal values. Also note that while $G_k$ is a formula in $k$-CNF,
$F$ is a formula in $(k+1)$-CNF (and its first part is in
$(k+1)$-CNF). At the first step, GUC satisfies a random literal
from a random clause of minimal size. With probability
$\frac{1}{3}$ this literal is $\n{x_1}$. In this case, our formula
becomes $G_k(x_{M+1},\ldots,x_{\lceil \frac{1+c_k}{c_k}M\rceil})$,
and the algorithm will have to make at least $poly(M)2^M$ choices
to eliminate all leaves of the assignment tree (\cite[Theorem
5.7]{Urqu}).

With probability $\frac{2}{3}$, GUC chooses another literal $l$ to
satisfy. Let $l=x_2$ (it does not matter which one we choose due
to symmetry).
\begin{multline*}
F[x_2]=(x_1\vee G(x_{M+1},\ldots,x_{\lceil
\frac{1+c_k}{c_k}M\rceil}))\wedge\\\wedge(\n{x_1}\vee
x_3\vee\n{x_4})\wedge(\n{x_1}\vee
x_4\vee\n{x_5})\wedge\cdots\wedge(\n{x_1}\vee
x_{M-1}\vee\n{x_M})\wedge(\n{x_1}\vee x_M)
\end{multline*}
The formula now has a 2-clause, and during the next step GUC will
either, with probability $\frac{1}{2}$, satisfy $\n{x_1}$, thus
creating a hard unsatisfiable instance, or satisfy $x_M$, and we
are left with
\begin{multline*}
F[x_2,x_M]=(x_1\vee G(x_{M+1},\ldots,x_{\lceil
\frac{1+c_k}{c_k}M\rceil}))\wedge\\\wedge(\n{x_1}\vee
x_3\vee\n{x_4})\wedge(\n{x_1}\vee
x_4\vee\n{x_5})\wedge\cdots\wedge(\n{x_1}\vee x_{M-1})
\end{multline*}
Only when there are no 3-clauses left, the last remaining literal
becomes a pure literal, and the last 2-clause is decided
automatically. It follows by easy induction that the probability
of setting $x_1=false$ (and forcing GUC to work for the time
$poly(M)2^M$) is $$P(x_1=false)=1-\frac{2}{3}2^{-M+3},$$ which
tends to $1$ exponentially fast as $M$ tends to $\infty$.

If we now denote by $N$ the total number of variables in the
formula, all of the above proves the following
\begin{thm}
For every $k\ge4$ there exists a set of satisfiable formulas
$F^k_N$ in $k$-CNF such that the modified GUC algorithm requires
to make at least $poly(N)2^{\frac{c_{k-1}}{1+c_{k-1}}N}$ choices
to find a satisfying assignment, and $F^k_N$ contains $N$
variables and no more than $aN$ clauses, where $a$ is a constant
not depending on $N$ and $c_k=O(1/k^{1/8})$.
\end{thm}

\section{Hard formulas for Randomized GUC}
\begin{figure}
\fbox{\parbox{12 cm}{
\parskip=-2mm
{\bf Making a choice with the Randomized GUC algorithm\\ Input:} A formula $F$ in CNF.\\
{\bf Method:}
\begin{enumerate}
\parskip=-1mm
\item $m:=min\{i:\ C^F_i\ne\emptyset\}$. \item If $m=1$ then
choose $C$ randomly from $C^F_1$ and set $l$ to the only literal
in $C$. \item Else if $PL(F)\ne\emptyset$, choose $l$ randomly
from $PL(F)$. \item Else choose $C$ randomly from $C^F_m$, then
choose $l$ randomly from $C\cup\n{C}$. \item Output $l$.
\end{enumerate}
}} \caption{One step of the Randomized GUC algorithm}\label{RGUC}
\end{figure}

It might seem that we succeeded with the GUC algorithm only
because of its highly determined behavior. The problem might be in
the necessary \emph{satisfying} a shortest clause. Our formula in
the preceding section ``tricks'' GUC into the wrong subtree
precisely because of this particular behavior. In this section, we
present a hard satisfiable instance for a modification of the GUC
algorithm, namely Randomized GUC algorithm. One step of this
algorithm is shown on Fig.\ref{RGUC}. It chooses a literal
randomly from the shortest clause, but also randomly chooses
whether to satisfy it. For example, if the shortest clause is
$a\vee b$, Randomized GUC could choose any literal of the set
$\{a,b,\n{a},\n{b}\}$.

Randomized GUC would break the example in the preceding section.
Indeed, on the very first step it will have a chance of
$\frac{1}{6}$ to set $x_1=true$, thus reducing the formula to a
very simple one. Therefore, by restarting Randomized GUC we can
achieve arbitrarily high probability of success.

Let us consider the following formula (denoting
$G:=G_k(x_{M+1},\ldots,x_{\lceil \frac{3c_k+2}{3c_k}M\rceil})$ and
assuming $3\mid M$ without loss of generality):
\begin{multline*}
F=(x_1\vee G)\wedge(x_2\vee G)\wedge\ldots\wedge(x_M\vee G)\wedge\\
\wedge(x_1\vee\n{x_2}\vee\n{x_3})\wedge(x_2\vee\n{x_3}\vee\n{x_1})\wedge(x_3\vee\n{x_1}\vee\n{x_2})\wedge\\
\wedge(x_4\vee\n{x_5}\vee\n{x_6})\wedge(x_5\vee\n{x_6}\vee\n{x_4})\wedge(x_6\vee\n{x_4}\vee\n{x_5})\wedge\ldots\\
\ldots\wedge(x_{M-2}\vee\n{x_{M-1}}\vee\n{x_M})\wedge(x_{M-1}\vee\n{x_M}\vee\n{x_{M-2}})\wedge(x_{M-1}\vee\n{x_M}\vee\n{x_{M-2}}).
\end{multline*}

As in the case described above, an assignment satisfies $F$ if and
only if it sets the variables $x_1,x_2,\ldots,x_M$ to $true$.

The Randomized GUC algorithm will first choose a random clause
among the shortest ones, that is, among the second part of our
formula, and then a random literal from the chosen clause. Since
all literals appear symmetrically, in fact it chooses a random
literal among $x_1,\ldots,x_M,\n{x_1},\ldots,\n{x_M}$ to satisfy
with equal probabilities. First note that if it chooses any of the
negative literals, $F[\n{x_i}]$ would contain $G$ as an
independent subformula. It would take exponentially long for
Randomized GUC to prove its unsatisfiability, since a
contradiction can be reached only in this subformula (it is easy
to see that the rest of the clusters of three clauses cannot be
reduced to an empty clause).  So, with probability $\frac{1}{2}$,
the desired result is achieved. Suppose it chooses $x_1$ (without
loss of generality, because the formula is symmetrical with
respect to the first $M$ variables). The formula now contains two
2-clauses, $(\n{x_2}\vee x_3)$ and $(\n{x_3}\vee x_2)$. The
algorithm now has to make a free choice with probability
$\frac{1}{2}$ of success (that is, choosing $x_2$ or $x_3$ rather
than a negation). If it succeeds, it gets a unit clause on the
next step and chooses a value for the remaining variable
correctly.

In short, every cluster of three 3-clauses with similar variables
has a probability of $\frac{1}{4}$ of setting the correct values
for its variables, and the algorithm considers these clusters one
at a time, one after another. Therefore, the overall probability
of success is $$P(\forall i:\\ 1\le i\le M\
x_i=true)=\frac{1}{2}2^{-\frac{2}{3}M}.$$ And in case of failure,
the time Randomized GUC will require to prove the unsatisfiability
of $G$ is $poly(M)2^{\frac{2}{3}M}$.  All of the above proves the
following

\begin{thm}
For every $k\ge4$ there exists a set of satisfiable formulas
$F^k_N$ in $k$-CNF such that the Randomized GUC algorithm requires
to make at least $poly(N)2^{\frac{2c_{k-1}}{2+3c_{k-1}}N}$ choices
to find a satisfying assignment, and $F^k_N$ contains $N$
variables and no more than $aN^2$ clauses, where $a$ is a constant
not depending on $N$ and $c_k=O(1/k^{1/8})$.
\end{thm}

\section{Further work}

In this paper we proved an exponential lower bound for satisfiable
formulas for two DPLL-type algorithms. However, ``hard'' formulas
for the Randomized GUC algorithm turned out to have quadratic
relationship between the number of clauses and the number of
variables. It would be interesting to construct similar
linear-sized formulas.

Also, apart from the unit clause and pure literal principles, a
number of other heuristics is used in modern DPLL-type SAT
solvers. Such heuristics include the resolution rule,
``black-and-white literals'' principle etc. (for more information
see \cite{DHIV}). Similar bounds are still to be proven for
algorithms employing these heuristics.

\section*{Acknowledgements} I would like to thank Edward A. Hirsch, who suggested the problem and supervised my work.

\end{document}